\pdfoutput=1

\documentclass[]{kuaishou}

\usepackage[toc,page,header]{appendix}
\usepackage{array}
\usepackage{siunitx,array}
\usepackage{makecell}
\usepackage{framed}
\usepackage{colortbl}

\usepackage{bbding}
\usepackage{amsmath}
\usepackage{amsfonts}
\usepackage[colorinlistoftodos]{todonotes}
\usepackage{longtable}
\usepackage{hhline}
\usepackage{fancyvrb}
\usepackage{float}
\usepackage{fvextra}
\usepackage{CJKutf8}
\usepackage{multicol}
\usepackage{tablefootnote}
\usepackage{threeparttable}
\usepackage{tabularx}
\usepackage{mdframed}
\usepackage[usestackEOL]{stackengine}
\usepackage[numbers]{natbib}
\newcommand{\commentout}[1]{}
\renewcommand{\paragraph}[1]
{\noindent\textbf{#1.}\hspace*{1em}}
\usepackage{enumitem}
\setlist[itemize]{leftmargin=15pt}
\usepackage{adjustbox}
\usepackage{arydshln}
\usepackage{pifont}
\usepackage{pifont}
\usepackage{xcolor}

\usepackage{booktabs}
\usepackage{makecell}
\usepackage{adjustbox}
\usepackage{array}
\usepackage[table]{xcolor}
\usepackage{pifont}

\RequirePackage{xspace}
\makeatletter
\DeclareRobustCommand\onedot{\futurelet\@let@token\@onedot}
\def\@onedot{\ifx\@let@token.\else.\null\fi\xspace}

\makeatother

\usepackage{array}
\usepackage[table]{xcolor}

\newcolumntype{Z}{>{\hspace*{\tabcolsep}}l<{\hspace*{\tabcolsep}}}

\title{Scaling Articulated Rationales for MLLM-based Recommendation}

\author{
\raisebox{-0.10em}{\includegraphics[height=0.94em]{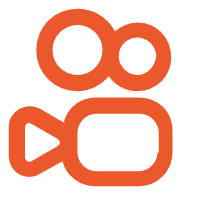}}
\hspace{-0.20em}
Kuaishou Technology
}

\vspace{-11pt}

\contribution{See \hyperref[sec:contributions]{Contributions} section for a full author list.}

\abstract{
Users' explanations of why they like or dislike content provide
explicit preference information that behavioral feedback and
content descriptions alone do not fully capture.
However, these articulated user rationales (AURs) are sparse,
uneven in quality, and limited in coverage, making them difficult
to use in industrial recommendation.
We present \textbf{SARA}
(\textbf{S}caling \textbf{A}rticulated \textbf{Ra}tionales),
a framework that transforms sparse AURs into recommendation
signals at scale.
SARA first collects and curates questionnaire responses into
\textbf{SARA-HQ}, an author-centric dataset that grounds model
alignment in users' stated preferences.
Using supervised fine-tuning and Quality-Refining DPO,
\textbf{SARA-7B} learns to generate positive and negative
rationales from multimodal author context, expanding coverage
from $86{,}564$ questionnaire-covered authors to the full
$10$M-author space.
\textbf{SARA-Ranker} then incorporates these rationales into
user--author interaction modeling and negative-feedback
history modeling.
Evaluation on unseen authors shows that SARA-7B produces
more specific, relevant, and grounded rationales than strong
general-purpose MLLM baselines.
On top of an established multimodal ranking system,
separate online A/B tests show that positive-rationale
integration increases watch time by $0.99\%$, while
negative-rationale integration reduces Hate feedback
by $8.16\%$.
With daily refresh and more than $30$ days of production
deployment, SARA demonstrates a practical path for using
MLLMs to scale sparse human explanations into effective
recommendation features.
\bigskip

\textbf{Date:} Sep 10, 2026

\textbf{Correspondence:} hk.xiao.me@gmail.com

}

\begin{document}

\maketitle
\vspace{-10pt}

\vspace{-4pt}

\newpage
\tableofcontents
\newpage

\section{Introduction}
\label{sec:intro}

Modern recommendation systems (RS) primarily infer user preferences from behavioral feedback, including clicks, watch time, skips, and explicit negative actions, together with user profiles and interaction histories~\cite{DBLP:reference/sp/RicciRS15,DBLP:journals/csur/ZhangYST19}.
\
These signals are abundant and easy to collect, but they mainly reveal \emph{what} users do rather than \emph{why} they make a decision.
The same observed behavior can arise from different motivations, while a skip or dislike alone does not specify which aspect of the content should be avoided.
Consequently, behavior-only supervision provides limited support for modeling fine-grained, transferable user preferences.

To bridge the semantic gap left by behavioral feedback, modern recommendation systems increasingly incorporate multi-modal signals to enrich their representations of users and items.
Visual signals capture objects, scenes, appearances, and activities, while audio signals convey speech, music, vocal characteristics, and acoustic context.
Textual signals provide complementary high-level semantics and user-generated opinions: hashtags help group related content~\cite{DBLP:conf/bigdataservice/BalineniA23}; news articles, headlines, and item descriptions characterize content semantics~\cite{DBLP:journals/ipm/KarimiJJ18}; and reviews and comments provide opinionated evidence~\cite{DBLP:journals/csur/HasanRDHR26}.
Recent MLLMs offer a unified interface for understanding and integrating these heterogeneous sources of information.

Nevertheless, richer multimodal representations do not necessarily reveal why a particular user likes or dislikes an item.
Visual and audio signals primarily describe what appears or occurs in the content, while hashtags and metadata summarize its topics and attributes.
Reviews and comments may contain subjective opinions, but they are produced under unconstrained self-expression and are not necessarily associated with a specific preference decision.
This missing connection between multimodal content understanding and user preference motivates \textbf{articulated user rationales (AURs)}, which explicitly capture why a user likes or dislikes an item.

Yet AURs have rarely been used as scalable recommendation signals, primarily due to three challenges:

\vspace{5pt}
\noindent$\bullet$ \textbf{R1. Native sparsity:} After consuming an item, the most natural user behavior is to scroll away, skip, or remain silent. 
\
Explicitly articulating why one likes or dislikes an item is an extremely low-probability event, typically with a coverage rate below $1\%$~\cite{nielsen2006participation}.

\noindent$\bullet$ \textbf{R2. Generally low expression quality:} AURs entered through UI interfaces are often colloquial and fragmented, exhibit low information density, and are affected by emotional bias introduced through voluntary self-reporting~\cite{douglas2023data}.

\noindent$\bullet$ \textbf{R3. Speculative generation by general-purpose MLLMs:}
Extending AUR coverage with general-purpose MLLMs is unreliable.
Even when enhanced by behavioral signals, MLLMs observe preference outcomes rather than the underlying reasons, and may therefore generate plausible but post-hoc rationalizations that are not grounded in users' articulated preferences~\cite{DBLP:conf/recsys/BaoZZWF023,DBLP:conf/sigir/LiaoL0WYW024}.

\begin{figure}[!ht]
    \centering
    \includegraphics[width=0.95\linewidth]{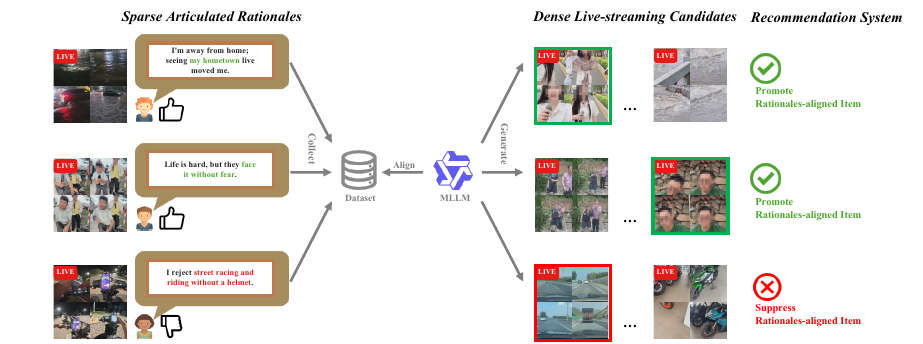}
    \caption{Overview of SARA.}
    \label{fig:intro}
\end{figure}

In this work, we present \textbf{SARA} (\textbf{S}caling \textbf{A}rticulated \textbf{Ra}tionales), the first industrial framework that scales sparse AURs into production-grade recommendation signals and demonstrates their practical value in a deployed system.

As illustrated in Fig.~\ref{fig:intro}, SARA connects articulated reasons to downstream recommendation through a standard \emph{collect--scale--integrate} pipeline: it collects and curates native AURs, aligns an MLLM with this articulated-preference supervision, applies the aligned model to generate rationales for $10 \mathrm{M}$ authors, and integrates the resulting rationale signals into the ranking model to enhance recommendation.

Specifically, SARA comprises the following key components:
\vspace{5pt}

\noindent$\diamond$ \textbf{Data Engine.} 
\
To address data sparsity and quality, we design a data engine that continuously elicits AURs from $240 \mathrm{M}$ Kuaishou Live users and applies post-collection curation.
\
By August 2026, the data engine had curated 187,532 refined rationales, forming \textbf{SARA-HQ}.

\noindent$\diamond$ \textbf{Rationale-Oriented Alignment.}
Based on SARA-HQ, we adapt a general-purpose MLLM into a rationale generator through a two-stage alignment pipeline: large-scale SFT for basic alignment, followed by Quality-Refining DPO for quality enhancement.
We term the resulting model \textbf{SARA-7B}.

\noindent$\diamond$ \textbf{Rationale-as-Feature Integration.}
Based on SARA-7B, we generate positive and negative rationales for $10\mathrm{M}$ Kuaishou Live authors and integrate them into the production ranking model through a two-branch architecture: rationale-aware interaction modeling for positive preference matching, and rejection-memory modeling for generalizing negative feedback across semantically related authors.
We term the resulting ranking model \textbf{SARA-Ranker}.

SARA promotes sparse articulated user rationales to first-class input signals for MLLM-based recommendation systems and delivers measurable online gains in real-world industrial deployment.
\
Beyond extensively explored multi-modal signals, articulated rationales are introduced into recommendation systems for the first time at industrial scale with daily-refresh deployment, establishing a new category of textual signals consumable by industrial recommendation systems.
\
We hope this work can provide useful insights to the community on the exploration and utilization of scalable, high-value signals that bring tangible benefits to recommendation systems.

\providecommand{\cmark}{\textcolor{green!50!black}{\ding{51}}}
\providecommand{\xmark}{\textcolor{red!75!black}{\ding{55}}}

\section{Data Engine}

\subsection{Overview}
\begin{figure}[H]
    \centering
    \includegraphics[width=1.00\linewidth]{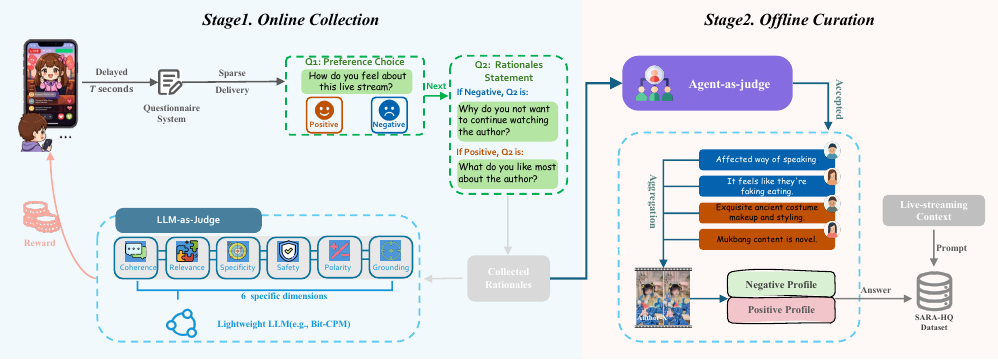}
    \caption{Overview of the SARA data engine.}
    \label{fig:data-collection}
\end{figure}

The data engine consists of two components: online data collection and post-curation. 
\
The online collection pipeline is built to continuously gather large-scale, native AURs, while the post-curation stage focuses on distilling high-value rationale signals from the naive collections.
\
The two components are described in detail below.
\
Figure~\ref{fig:data-collection} provides an overview of the complete
quantity--quality data engine before detailing its online collection and
offline curation stages.

\subsection{Online Collection}

In practice, the data collection process is conducted through a questionnaire-based sending and receiving system, which is ubiquitously deployed across the Kuaishou live-streaming ecosystem and is indiscriminately exposed to all users. 
\
The questionnaire content consists of two steps: 

\vspace{5pt}
\noindent$\diamond$ \textbf{Preference Choice}. Based on the current live-streaming context, users select the polarity preference from \textbf{Positive} and \textbf{Negative}. 

\noindent$\diamond$ \textbf{Articulated Rationales Statement}. Conditioned on the selected preference, the system dynamically issues the next open-ended follow-up question: if the user selects Positive, the system asks, \textbf{“What do you like most about this live streaming?
”}; if the user selects Negative, the system asks, \textbf{“Why do you feel negative to this live streaming?}” Herein, users state their articulated polarity rationales for live streaming context.

The questionnaire above defines what preference feedback is collected. At production scale, the collection process must further balance data quantity with response quality. We therefore introduce the following mechanisms to improve both aspects jointly.

\vspace{5pt}

\noindent$\diamond$ \textbf{Delayed Triggering Mechanism.}
The questionnaire is issued only after a user has continuously watched the live stream for $\mathcal{T}$ seconds. If $\mathcal{T}$ is too short, users may not yet have formed a stable perception of the streaming content, resulting in arbitrary or ambiguous responses. Conversely, if $\mathcal{T}$ is too long, users with negative attitudes may leave the stream before the questionnaire is triggered, leading to missed collection opportunities. Balancing these trade-offs, we empirically set $\mathcal{T}=10$ seconds as an engineering compromise between response quality and collection coverage.

\noindent$\diamond$ \textbf{Sparse Questionnaire Delivery Mechanism.} 
Industrial practice suggests that overly frequent questionnaire exposure can induce survey fatigue, thereby reducing users’ willingness to respond. Moreover, responses collected under fatigued conditions provide limited marginal value to the data engine, ultimately reducing the overall efficiency of data collection.  
To mitigate survey fatigue, questionnaires are delivered to users with a small probability $\mathcal{P}$. The value of $\mathcal{P}$ is determined through small-scale A/B testing: after evaluating candidate probabilities ranging from $1\%$ to $20\%$, we select $5\%$ as the baseline probability that balances collection scale and user fatigue.
Therefore, we dynamically adjust the delivery probability according to the recent response rate:

\begin{equation}
\mathcal{P} = Min \left( Max \left( p_0 \times \frac{r_{\text{obs}}}{r_{\text{target}}},\; p_{\text{min}} \right),\; p_{\text{max}} \right)
\end{equation}

where $p_0=5\%$ denotes the baseline probability,
$r_{\mathrm{obs}}$ is the observed response rate within a sliding window (the past 5 minutes),
$r_{\mathrm{target}}=12\%$ is the target response rate,
$p_{\min}=1\%$,
and $p_{\max}=20\%$.
This dynamic adjustment mechanism increases questionnaire exposure when users respond actively and decreases exposure when response willingness declines, while constraining the probability within a reasonable range. As a result, data collection is concentrated during periods of higher user engagement while avoiding excessive interruption.

\noindent$\diamond$ \textbf{Threshold-based Incentive Mechanism.}
We deploy a lightweight online LLM judge, eg. BitCPM4-0.5B~\cite{DBLP:journals/corr/abs-2506-07900}, to evaluate each response in real time against the available live-streaming context.
Following the quality protocol described in \S\ref{sec:rationale-evaluation}, the judge assesses six dimensions: \textit{coherence}, \textit{relevance}, \textit{specificity}, \textit{safety}, \textit{polarity consistency}, and \textit{grounding}.
It also produces a holistic quality score $\widehat{G}_{\mathrm{online}}\in\{1,2,3,4\}$ within the same inference pass.
A response qualifies for a small \textit{KuaiCoin} reward if
\begin{equation}
\widehat{G}_{\mathrm{online}}(y)>2.
\end{equation}
This conditional positive reward mechanism encourages high-quality feedback and, in practice, effectively improves users' response willingness.

\vspace{5pt}
After offline optimization, the data collection pipeline was deployed in production. In the latest observation window from February 12 to August 30, 2026, it collected $1{,}915{,}718$ candidate AURs over 185 active days, or $10.4\text{K}$ responses per day on average.

\subsection{Post Curation}
\label{sec:post_curation}

Post-curation converts the $1.92\mathrm{M}$ collected candidates into quality-controlled supervision.
The online LLM judge is used exclusively for real-time incentive allocation, whereas offline quality filtering is performed directly on the raw responses by the higher-fidelity Agent Judge, without using the online scores as filtering inputs.
We describe its evaluation framework and validation in \S\ref{sec:rationale-evaluation}.
Briefly, six specialists assess \textit{coherence}, \textit{relevance}, \textit{specificity}, \textit{safety}, \textit{polarity consistency}, and \textit{grounding}, respectively providing dimension-level score.
A Senior Reviewer consolidates these assessments and produces a holistic quality score $\widehat{G}_{\mathrm{agent}}$.
We retain a rationale when $\widehat{G}_{\mathrm{agent}}>2$, with the additional constraint that a critical defect in relevance, safety, polarity consistency, or grounding cannot be offset by strong performance on the other dimensions.
This procedure retains $187{,}532$ high-quality rationales, corresponding to an overall offline yield of $9.8\%$.
\
We subsequently organize the retained rationales into author-centric supervision.
\
The UA matrix constructed from 187,532 high-quality rationales has dimensions of 141,461 × 86,564.
The collected user--author questionnaire matrix is highly sparse and asymmetric: each responding user contributes only approximately $1.33$ questionnaires on average, whereas each covered author receives approximately $ 2.17 $ responses from different users.
The limited per-user coverage is insufficient to support reliable user-level preference understanding.
In contrast, aggregating responses from multiple users for the same author reveals recurring factors that attract or repel viewers.
We therefore group the verified rationales by author and preference polarity, converting sparse individual feedback into more informative author-level profiles of positive and negative preference factors, ultimately constructing $86,564$ author-centric profiles with polarity-aware characteristics.
\
Individual questionnaire responses are typically short and fragmentary, with $82\%$ containing no more than ten Chinese characters.
Despite their brevity, these responses often express a concrete and self-contained preference factor, while different users frequently describe the same factor using colloquial or synonymous expressions.
This combination of semantic atomicity and surface-form variation makes the responses well suited for tag-based normalization.
We therefore convert the rationales associated with each author and polarity into atomic semantic tags, consolidating synonymous expressions, removing redundancy, and retaining only tags grounded in the verified responses.

\paragraph{Data Construction}
Following the multimodal instruction-tuning formulation of
LiViBench~\cite{wang2026livibench}, we organize interactive livestream
evidence into instruction--response instances for subsequent alignment.
We adapt this construction from generic livestream question answering
to preference-conditioned rationale generation: rather than synthesizing
QA targets, we use articulated reasons distilled from real-user
questionnaires as supervision.

Each training instance is represented as $(\mathcal{X}_p,r_p)$, where
\begin{equation}
\mathcal{X}_p =
\left(
V,
T_{\mathrm{meta}},
T_{\mathrm{asr}},
T_{\mathrm{comment}},
I_p
\right).
\end{equation}
Here, $V$ contains sampled livestream frames, while
$T_{\mathrm{meta}}$, $T_{\mathrm{asr}}$, and
$T_{\mathrm{comment}}$ denote the stream metadata, 
ASR transcript, and comment of viewer, respectively. The
polarity-conditioned instruction $I_p$ asks the model to generate
positive rationales when $p=+$ and negative rationales when $p=-$.
The ground-truth answer $r_p$ is distilled and normalized from the
corresponding high-quality original questionnaire responses retained
by the data engine. 
The resulting collection of $187,532$ quality-filtered,
polarity-aware rationale instances constitutes our final dataset,
\textbf{SARA-HQ}.

\begin{figure}[H]
    \centering
    \includegraphics[width=\textwidth]{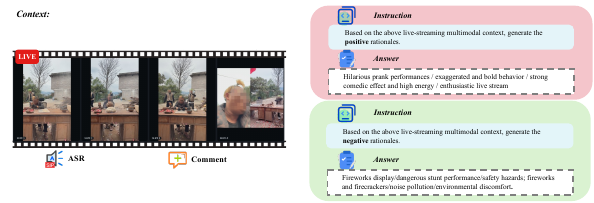}
    \caption{
    Illustration of SARA-HQ data construction. 
    }
    \label{fig:sara-data-case}
\end{figure}

Figure~\ref{fig:sara-data-case} provides a concrete example. The
sampled frames show a firecracker performance in a rural courtyard;
the ASR captures loud promotional shouting and references to larger
explosions, while the comments reflect reactions to the host's
appearance and performance. Given this shared evidence, the positive
instruction corresponds to rationales centered on humor, energy, and
interaction, whereas the negative instruction corresponds to rationales
concerning dangerous behavior, safety hazards, noise, and pollution.
Thus, the case illustrates how one aligned livestream context supports controllable,
polarity-specific rationale generation without leaking questionnaire
content into the context.

\subsection{Dataset Statistics and Analysis}
\label{sec:dataset-analysis}

\paragraph{Pipeline Statistics.}
Table~\ref{tab:data-engine-validation} reports the latest snapshot of the data engine. From February 12 to August 30, 2026, online collection accumulated $1{,}915{,}718$ candidate AURs. Applying the offline criterion $\widehat{G}_{\mathrm{agent}}>2$ retains $187{,}532$ rationales, giving an overall high-quality (HQ) yield of $9.8\%$. We call this quality-controlled supervision pool \textbf{SARA-HQ}, using a size-agnostic name that remains valid as the production collection continues to grow.

\begin{table*}[!ht]
\centering
\small
\caption{Latest statistics of the SARA data engine. HQ denotes responses assigned holistic levels 3--4 by the offline Agent Judge. The categorized subset contains retained rationales with valid first- and second-level semantic categories.}
\label{tab:data-engine-validation}
\setlength{\tabcolsep}{3pt}
\begin{tabular}{l|c|c|c}
\toprule
\textbf{Stage / Subset} & \textbf{Candidate AURs} & \textbf{HQ AURs (Yield)} & \textbf{Role} \\
\midrule
Positive & $1{,}584{,}866$ & $107{,}540$ ($6.8\%$) & Polarity-specific supervision \\
Negative & $330{,}852$ & $79{,}992$ ($24.2\%$) & Polarity-specific supervision \\
All collected & $1{,}915{,}718$ & $187{,}532$ ($9.8\%$) & Offline quality filtering \\
Categorized HQ subset & -- & $184{,}142$ ($98.2\%$ of HQ) & Coverage analysis \\
\textbf{SARA-HQ} & -- & \textbf{$187{,}532$} & Final training supervision \\
\bottomrule
\end{tabular}
\end{table*}

\begin{figure}[H]
    \centering
    \includegraphics[width=\textwidth]{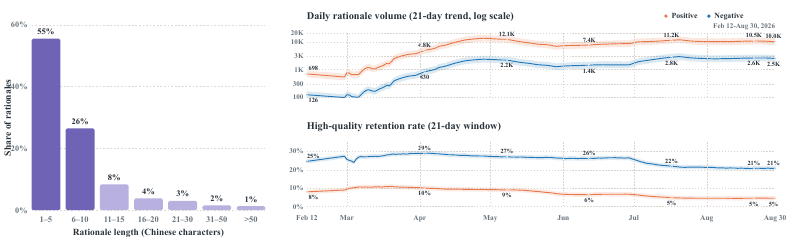}
    \caption{Scale and quality characteristics of SARA-HQ. Left: distribution of rationale length in Chinese characters. Right: 21-day rolling candidate volume and offline HQ yield for positive and negative AURs from February 12 to August 30, 2026. Shaded regions indicate local variability around each trend.}
    \label{fig:scale-quality-analysis}
\end{figure}

Figure~\ref{fig:scale-quality-analysis} complements the aggregate table with granularity and temporal evidence. Among $6{,}134$ segmented questionnaire rationales used for the length analysis, $82\%$ contain no more than ten Chinese characters (median $5$, mean $8.25$). This confirms that individual AURs are informative but often too terse to serve as stable author descriptions in isolation, motivating author-centric aggregation and semantic normalization. The temporal panel shows that collection scales to roughly $10\text{K}$ positive and $2.5\text{K}$ negative candidates per day by the end of the observation window. At the same time, the 21-day HQ yield settles near $5\%$ for positive and $21\%$ for negative candidates. The smaller negative pool is more defect-focused and therefore passes the strict offline criterion more frequently, whereas abundant positive feedback contains more generic praise that is removed during curation.

\paragraph{Comparison with Public Livestreaming Datasets.}
Table~\ref{tab:dataset-comparison} compares SARA-HQ with representative publicly released datasets for live-streaming recommendation and understanding. Because their released units differ substantially, we report both scale and supervision source rather than treating clips, streams, interactions, questionnaires, and rationales as directly interchangeable. We further distinguish whether the effective supervision is human-derived, preference-bearing, polarity-aware, open-ended, and reason-level. Here, \emph{open-ended} means that free-form text is directly used as a training or evaluation target, whereas \emph{reason-level} more specifically means that an individual natural-language explanation of user preference constitutes the supervision unit.

\begin{table*}[!ht]
\centering
\small

\caption{
Comparison with representative public livestream datasets and
benchmarks. ``Protocol'' indicates whether a dataset supports
online streaming inference or assumes access to the complete
offline context.
``Human'' indicates that the supervision is derived from real-user
feedback or human annotation/verification. ``Preference'' distinguishes
explicit feedback from implicit behavioral signals. ``Polarity''
indicates explicit positive and negative preference labels.
``Open-ended'' indicates whether free-form text is directly used as
training or evaluation supervision. ``Reason-level'' indicates that
individual natural-language explanations of user preference serve as
direct supervision targets. Although KuaiLive-M3 collects open-ended
responses in Q2, its benchmark experiments use only the categorical
preference labels from Q1.
}
\label{tab:dataset-comparison}

\setlength{\tabcolsep}{2.1pt}
\renewcommand{\arraystretch}{1.12}

\newsavebox{\SARAHQrowbox}
\newcommand{\SARAHQrow}{%
\textbf{SARA-HQ}
& \textbf{Ours}
& Offline
& \makecell[l]{
    $187.5$K quality-controlled rationales\\
    $18$ domains, $141$ second-level categories
}
& \makecell[l]{
    Agent-judged real-user\\questionnaires
}
& \cmark
& Explicit
& \cmark
& \cmark
& \cmark
}
\sbox{\SARAHQrowbox}{%
  \begin{tabular}{lccllccccc}
    \SARAHQrow \\
  \end{tabular}%
}

\begin{adjustbox}{max width=\textwidth}
\begin{tabular}{lccllccccc}
\toprule

\textbf{Dataset}
& \textbf{Pub.}
& \textbf{Protocol}
& \textbf{Scale}
& \textbf{Supervision source}
& \textbf{Human}
& \textbf{Preference}
& \textbf{Polarity}
& \makecell[c]{\textbf{Open-}\\\textbf{ended}}
& \makecell[c]{\textbf{Reason-}\\\textbf{level}} \\

\midrule

\rowcolor{gray!12}
\multicolumn{10}{l}{
    \textit{Livestream recommendation and interaction datasets}
} \\
\addlinespace[1pt]

LiveRec~\cite{rappaz2021liverec}
& RecSys'21
& Offline
& \makecell[l]{
    $15.5$M users, $465$K streamers\\
    $474.7$M interactions, $43$ days
}
& \makecell[l]{
    Chat-derived user--streamer\\
    co-presence logs
}
& \cmark
& Implicit
& \xmark
& \xmark
& \xmark \\
\addlinespace[2pt]

KuaiLive~\cite{qu2026kuailive}
& SIGIR'26
& Offline
& \makecell[l]{
    $23.8$K users, $452.6$K streamers\\
    $5.36$M interactions, $21$ days
}
& \makecell[l]{
    Timestamped multi-behavior\\
    interaction logs
}
& \cmark
& Implicit
& \xmark
& \xmark
& \xmark \\
\addlinespace[2pt]

KuaiLive-M3~\cite{guo2026kuailivem3}
& arXiv'26
& Offline
& \makecell[l]{
    $21.9$K users; $35$M live and\\
    $111$M short-video interactions\\
    $88$M embeddings; $25.4$K questionnaires
}
& \makecell[l]{
    Multi-domain behavior logs\\
    and questionnaire responses
}
& \cmark
& \makecell[c]{Explicit\\ \& implicit}
& \cmark
& \xmark
& \xmark \\

\midrule

\rowcolor{gray!12}
\multicolumn{10}{l}{
    \textit{Livestream understanding datasets and benchmarks}
} \\
\addlinespace[1pt]

LiveCC~\cite{chen2025livecc}
& CVPR'25
& Streaming
& \makecell[l]{
    $5$M pretraining clips\\
    $526$K SFT clips
}
& \makecell[l]{
    Timestamp-aligned CC/ASR
}
& \cmark
& --
& \xmark
& \cmark
& \xmark \\
\addlinespace[2pt]

OmniStar~\cite{yang2025livestar}
& NeurIPS'25
& Streaming
& \makecell[l]{
    $20{,}137$ video streams\\
    $19{,}137$ train, $1{,}000$ evaluation
}
& \makecell[l]{
    Expert-annotated temporal\\
    captions and streaming QA
}
& \cmark
& --
& \xmark
& \cmark
& \xmark \\
\addlinespace[2pt]

LiViBench~\cite{wang2026livibench}
& AAAI'26
& Offline
& \makecell[l]{
    $3{,}168$ videos, $3{,}175$ MCQs\\
    $49.1$K instruction-tuning samples
}
& \makecell[l]{
    Model-proposed, human-verified\\
    multimodal MCQs
}
& \cmark
& --
& \xmark
& \xmark
& \xmark \\
\addlinespace[2pt]

LiveLongBench~\cite{wu2026livelongbench}
& \makecell[c]{ACL Findings\\'26}
& Offline
& \makecell[l]{
    $967$ evaluation instances\\
    $\sim97$K tokens per context
}
& \makecell[l]{
    Human-labeled QA based on\\
    manually corrected ASR
}
& \cmark
& --
& \xmark
& \cmark
& \xmark \\

\addlinespace[2pt]
\noalign{%
  \begingroup
  \color{gray!12}%
  \hrule height\dimexpr\ht\SARAHQrowbox+\dp\SARAHQrowbox\relax
  \kern-\dimexpr\ht\SARAHQrowbox+\dp\SARAHQrowbox\relax
  \endgroup
}
\SARAHQrow \\

\bottomrule
\end{tabular}
\end{adjustbox}
\end{table*}

Existing public resources fall into two broad groups. Recommendation datasets provide human-derived preference evidence, but Twitch/LiveRec and KuaiLive express it only through implicit co-presence or interaction logs~\cite{rappaz2021liverec,qu2026kuailive}. KuaiLive-M3 further combines implicit behavior with explicit questionnaire feedback and maps its three Q1 choices to positive and negative preference labels~\cite{guo2026kuailivem3}. Although it also collects an open-ended Q2 response, the released benchmark uses only the categorical Q1 labels. Free-form reasons are therefore neither the effective experimental target nor the supervision unit in the public live-streaming recommendation datasets.

Livestream-understanding datasets instead provide substantially richer semantic supervision about stream content. LiveCC and OmniStar support streaming captioning or QA, LiViBench centers on human-verified multimodal MCQs, and LiveLongBench evaluates long-context QA over corrected speech transcripts~\cite{chen2025livecc,yang2025livestar,wang2026livibench,wu2026livelongbench}. Some of these targets are open-ended, but none is conditioned on user preference polarity or treats an articulated preference reason as the supervision unit. SARA-HQ occupies this missing intersection: it derives explicit positive and negative feedback from real users and converts their open-ended questionnaires into quality-controlled, reason-level targets grounded in video, ASR, metadata, and comments. The resulting supervision directly supports polarity-conditioned rationale generation and subsequent rationale-aware recommendation.

\begin{figure}[H]
    \centering
    \includegraphics[width=\textwidth]{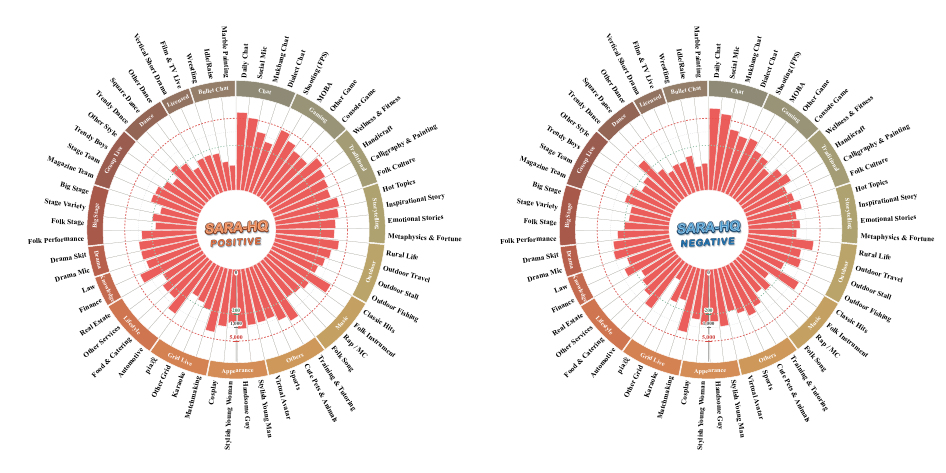}
    \caption{Semantic coverage of quality-filtered articulated user rationales. Panels show (a) positive and (b) negative rationales. The outer ring groups second-level categories by their first-level live-streaming domain, while radial bar length denotes rationale volume on a shared logarithmic scale. }
    \label{fig:category-polarity-coverage}
\end{figure}

\paragraph{Distribution and Semantic Coverage.}
Of the retained SARA-HQ rationales, $184{,}142$ receive valid semantic category assignments, covering 18 first-level live-streaming domains and 141 second-level categories. Fig.~\ref{fig:category-polarity-coverage} displays the four largest second-level categories in each eligible domain on a shared logarithmic scale. Among the 60 displayed categories, 46 positive and 37 negative categories contain more than 200 rationales, while two categories under each polarity exceed 5,000. The long tail therefore remains substantial after strict quality filtering rather than collapsing into a few head categories.

\begin{figure}[H]
    \centering
    \includegraphics[width=0.92\linewidth]{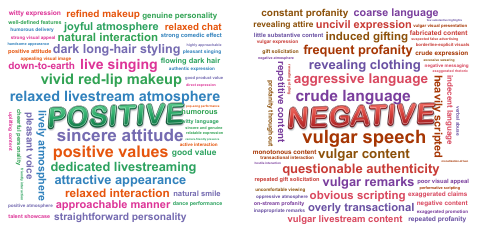}
    \caption{Semantic-tag word clouds for positive and negative SARA-HQ rationales. 
    }
    \label{fig:wordcloud}
\end{figure}

Although negative candidates are less frequent overall, the retained negative rationales remain broadly distributed and tend to express more specific defects. Fig.~\ref{fig:wordcloud} shows the word cloud of selected semantic tags: positive feedback emphasizes interaction, atmosphere, performance, and authenticity, whereas negative feedback concentrates on language, scripting, authenticity, and transactional behavior.

\section{Rationale-Oriented Alignment}
\label{sec:alignment}
\subsection{Overview}

The SARA-HQ dataset covers only $86,564$ authors, which is far less than the total number of full live-streaming authors. Therefore, we align the general MLLM Qwen2.5-VL-7B~\cite{DBLP:journals/corr/abs-2502-13923} with SARA-HQ to further generate rationales for full live streaming authors. 
\
The entire alignment process consists of two stage: \textbf{Large-Scale SFT} and \textbf{Quality-Refining DPO}, as detailed below, with the final model termed as SARA-7B.

\subsection{Large-Scale SFT}
Same as described in \S\ref{sec:post_curation}, we define the SFT task such that, given a multimodal context $\mathcal{X}=(V,\allowbreak T_{\text{meta}},\allowbreak T_{\text{comment}},\allowbreak T_{\text{asr}},\allowbreak I)$, where $V$ denotes the live-streaming video, $T_{\text{meta}}$ its title and description, $T_{\text{asr}}$ the corresponding ASR transcript, $T_{\text{comment}}$ the associated comment stream, and $I$ the task instruction specifying a polarity $p\in\{+,-\}$, the model is required to generate the corresponding positive or negative rationale.
\
In doing so, we formulate polarity-conditioned rationale generation as a single-turn conditional generation problem, and then perform full-parameter supervised fine-tuning with the standard auto-regressive negative log-likelihood objective:
\begin{equation}
\mathcal{L}_{\text{SFT}}(\theta) = -\frac{1}{N} \sum_{t=1}^{N} \log \pi_\theta\!\left(r_t \mid r_{<t},\; \mathcal{X}\right),
\end{equation}
where $N$ is the length of target rationale text, $r_t$ is the $t$-th token, and $r_{<t}=(r_1,\dots,r_{t-1})$.
\
At this stage, the general MLLM comprehensively learned from large-scale SARA-HQ instances how to articulate author-centric rationales by analyzing multimodal livestream context conditioned on the queried polarity.

\subsection{Quality-Refining DPO}

\begin{figure}[!htbp]
    \centering
    \includegraphics[width=0.6\linewidth]{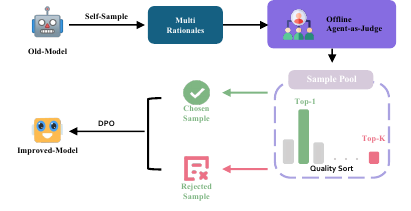}
    \caption{Overview of Quality-Refining DPO. 
    }
    \label{fig:dpo}
\end{figure}

To further improve the output quality of the SFT model, we draw inspiration from self-improvement techniques~\cite{DBLP:journals/corr/abs-2402-06457,DBLP:conf/nips/ZelikmanWMG22} and construct high-quality and low-quality preference pairs from responses sampled from the SFT model itself for subsequent DPO training. 
\
Moreover, this preference-pair construction process can be applied iteratively, enabling progressively improved models to be derived from their predecessors.
\
We refer to this algorithm as Quality-Refining DPO (QR-DPO), as illustrated in Fig.~\ref{fig:dpo} and detailed below.

\begin{figure}[!ht]
    \centering
    \includegraphics[width=0.60\linewidth]{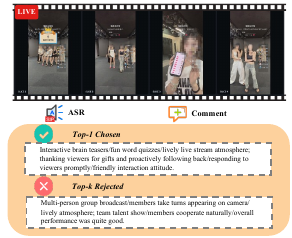}
    \caption{Qualitative examples of preference refinement in QR-DPO. 
    }
    \label{fig:qrdpo-cases}
\end{figure}

First, we sample $7\mathrm{K}$ popular authors from the SFT training set to construct the QR-DPO data.
\
For each associated training instance $\mathcal{X}$, we use the SFT-trained model $\pi_{\mathrm{SFT}}$ to sample $K$ candidate rationales $\{\hat{Y}^{(k)}\}_{k=1}^{K}$ at temperature $T$, forming the preference candidate set.
\
Inspired by agentic reward modeling~\cite{DBLP:conf/acl/0015QW000L25}, we employ the offline Agent Judge described in \S\ref{sec:rationale-evaluation} to evaluate these candidates under a unified six-dimensional rubric: coherence, relevance, specificity, safety, polarity consistency, and grounding. Dimension-specialized evaluators inspect each candidate against the multi-modal context and queried polarity, and a Senior Reviewer consolidates their scores and supporting evidence into a holistic quality assessment. These assessments guide candidate ranking, yielding an ordering \(\rho:\{1,\ldots,K\}\rightarrow\{1,\ldots,K\}\), where \(\rho(k)=1\) denotes the highest-ranked candidate. We select the rank-1 candidate as the chosen response \(y^{+}\) and the rank-\(K\) candidate as the rejected response \(y^{-}\), forming a preference pair \((y^{+},y^{-})\). This produces the preference dataset \(\mathcal{D}_{\mathrm{DPO}}=\{(X_i,y_i^{+},y_i^{-})\}_{i=1}^{N}\), with \(N=7\mathrm{K}\). We then optimize the model using the standard DPO loss, initializing the policy from \(\pi_{\mathrm{SFT}}\) and using the same checkpoint as the fixed reference policy \(\pi_{\mathrm{ref}}\):

\begin{subequations}
\begin{align}
r_\theta(y \mid X) &= \beta \log\frac{\pi_\theta(y \mid X)}{\pi_{\text{ref}}(y \mid X)} \label{eq:implicit_reward} \\
\mathcal{L}_{\text{DPO}}(\theta) &= -\,\mathbb{E}_{(X,y^+,y^-)\sim\mathcal{D}_{\text{DPO}}} \log\sigma\bigl( r_\theta(y^+ \mid X) - r_\theta(y^- \mid X) \bigr) \label{eq:dpo_loss}
\end{align}
\end{subequations}

Here, $\pi_\theta(y\mid X)$ denotes the probability assigned by the current model parameterized by $\theta$ to candidate $y$ given the context $X$, and $\pi_\text{ref}$ is the reference model, i.e., the SFT model $\pi_\text{SFT}$. The scaling factor $\beta$ controls the strength of preference optimization. The terms $r_\theta(y^+ \mid X)$  and $r_\theta(y^- \mid X)$ capture the log-probability ratios of the current model relative to the reference model on the chosen and rejected rationales, respectively. The sigmoid function $\sigma(\cdot)$ encourages the model to increase the probability of top-quality candidates while suppressing that of bottom-quality candidates, thereby refining and shifting the probability mass toward the top-quality region. 
\
Figure~\ref{fig:qrdpo-cases} illustrates the relative quality distinctions used to construct QR-DPO preference pairs.
Both candidates provide plausible rationales for the observed live-streaming content.
Compared with the rejected response, the chosen response identifies more concrete preference factors by connecting the brain-teaser activity and audience reactions to the perceived enjoyment and sincerity of the stream.
This relative advantage provides a finer-grained preference signal for refining rationale generation beyond the baseline quality already acquired through SFT.

\section{Rationale Quality Evaluation Framework}
\label{sec:rationale-evaluation}

\subsection{overview}
Reliable rationale-quality assessment is a common requirement across SARA.
During online collection, the system must determine whether a user response
deserves an incentive. During offline curation, the Data Engine must remove
noisy or unsupported articulated user rationales (AURs). During rationale
alignment, QR-DPO requires reliable preferences among self-sampled candidates,
and during final evaluation, generated rationales must be compared under a
consistent quality standard. We therefore establish a unified human-grounded
evaluation protocol for a candidate rationale $y$, conditioned on the
available livestream context $\mathcal{X}$ and queried polarity $p$. This follows the
growing use of rubric-guided LLMs as scalable evaluators of open-ended
generation~\cite{zheng2023judging,liu2023geval,kim2023prometheus}, while
adapting the evaluation criteria to articulated preference rationales in
livestream understanding task.

The protocol evaluates six dimensions: \emph{coherence}, \emph{relevance},
\emph{specificity}, \emph{safety}, \emph{polarity
consistency}, and \emph{grounding}. Each dimension is assessed on a
four-level scale, where scores $4$, $3$, $2$, and $1$ denote excellent,
acceptable, deficient, and poor performance, respectively. In addition to
these dimension-level scores, the protocol assigns a holistic-level score
$G\in\{1,2,3,4\}$ that reflects the final usability of the rationale under the
same four-level semantics. The holistic-level score is not the arithmetic
mean of the dimension-level scores. Instead, the dimensions are
non-compensatory: high
linguistic quality cannot offset a critical failure such as irrelevance,
contradiction of the queried polarity, an unsupported material claim, or a
safety violation. Detailed definitions and four-level score anchors for both
the dimension-level and holistic-level scores are provided in
Appendix~\ref{app:quality-rubric}.

\begin{figure*}[t]
    \centering
    \includegraphics[width=0.96\textwidth]{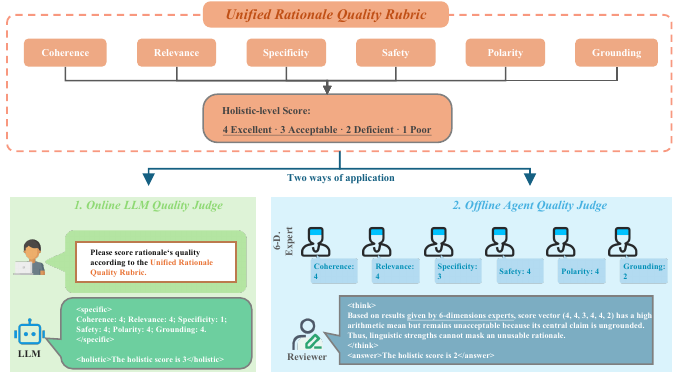}
    \caption{Overview of the rationale quality evaluation framework. A
    common six-dimensional, four-level quality standard supervises a
    single-pass online LLM and a process-oriented offline Agent Judge. The
    latter performs dimension-specialized inspection followed by Senior
    Review, where critical defects cannot be compensated for by high scores on
    other dimensions.}
    \label{fig:judge-framework}
\end{figure*}

We instantiate this common standard under two computational regimes. A
lightweight online LLM predicts all dimension-level scores and the
holistic-level score in a single pass, supporting latency-sensitive questionnaire
interaction. For high-fidelity offline assessment, an Agent-as-a-Judge
decomposes evaluation into dimension-specialized inspections followed by
Senior Review, following recent process-oriented and agentic
evaluation frameworks~\cite{zhuge2024agent,gou2025mind2web,DBLP:journals/corr/abs-2512-14503}.
Both evaluators are trained with the same human
annotations. The online judge is used for real-time incentive allocation,
whereas isolated instances of the agentic evaluator support offline data
curation, QR-DPO preference construction, and final model evaluation. An
overview is shown in Fig.~\ref{fig:judge-framework}. We next describe the
inference procedures of the two evaluator regimes, the human supervision used
to train them, and their alignment with held-out expert judgments.

\subsection{Online LLM Judge and Offline Agent Judge}
\label{sec:quality-evaluators}

\paragraph{Low-latency online LLM judge.}
The online judge is optimized for fixed-cost, low-latency assessment. Given
$(\mathcal{X},p,y)$, it predicts the six dimension-level scores and the
holistic-level score in a single pass:
\begin{equation}
(\widehat{\mathbf q},\widehat G)
=J_{\mathrm{online}}(\mathcal{X},p,y).
\end{equation}
At deployment, the system directly triggers the quality-aware incentive when
$\widehat G>2$. This outcome-oriented design
provides the efficient operating point required by online elicitation. For
high-fidelity offline assessment, we instead use the process-oriented agentic
evaluation described next.

\paragraph{Dimension-specialized evaluators.}
Single-pass evaluation must jointly attend to heterogeneous criteria and can
allow a strong global impression, such as fluency, to influence unrelated
judgments such as grounding. We instead instantiate one specialized evaluator
$E_d$ for each dimension $d$:
\begin{equation}
(q_d,e_d)=E_d(\mathcal{X},p,y;d),
\end{equation}
where $q_d$ is a four-level score and $e_d$ is a concise evidence report.

\paragraph{Senior Review.}
The Senior Reviewer receives the specialist reports
$\mathcal{S}(y)=\{(q_d,e_d)\}_{d=1}^{6}$ and produces a holistic-level score
$G\in\{1,2,3,4\}$:
\begin{equation}
G=R\!\left(\mathcal{S}(y)\right).
\end{equation}
The reviewer is not a score averager. It first performs critical-defect
detection over designated hard dimensions $\mathcal{D}_{\mathrm{hard}}$,
including relevance, safety, polarity consistency, and grounding:
\begin{equation}
\exists d\in\mathcal{D}_{\mathrm{hard}},\;q_d\leq2
\quad\Longrightarrow\quad G\leq2.
\label{eq:critical-gating}
\end{equation}
Only candidates without a critical defect are further distinguished between
acceptable ($G=3$) and excellent ($G=4$). For example, a score vector
$(4,4,4,4,4,1)$ has a high arithmetic mean but remains unacceptable because
its central claim is ungrounded. Thus, linguistic strengths cannot mask an
unusable rationale.

\subsection{Judge Corpus and Evaluator Training}
\label{sec:judge-corpus}

The two evaluator regimes above cannot be reliably instantiated with a
general-purpose LLM alone. Such models are not calibrated to articulated
preference rationales and may over-reward fluency while overlooking
livestream-specific context, the queried preference polarity, or unsupported
claims. We therefore convert the rubric into domain-specific human supervision
that covers both naturally occurring rationales and difficult quality
boundaries.

We construct the judge corpus from native AURs sampled from production logs,
rationales generated by diverse models and training checkpoints, and
controlled failures. Controlled failures are deliberately constructed or
transformed examples that exhibit predefined quality defects, enabling
targeted coverage of errors that may be sparse or entangled in naturally
occurring data. We target context mismatch, polarity reversal, generic content
description, and unsupported-detail injection. This mixture follows prior rubric-conditioned
judge training and challenging judge-evaluation protocols, which combine
diverse model outputs with targeted difficult
examples~\cite{kim2023prometheus,tan2024judgebench}. The controlled failures
serve only as candidate hard cases and domain experts verify the intended
transformation and annotate all affected quality dimensions rather than
treating automatically constructed labels as ground truth.

For each tuple $(\mathcal{X},p,y)$, domain experts provide six
dimension-level scores $\mathbf{q}^{*}=(q_1^{*},\ldots,q_6^{*})$ and a
holistic-level score $G^{*}$. For dimensions requiring contextual
verification, annotators additionally identify concise supporting or
contradicting evidence from the video, ASR, metadata, or comments. Samples are
split by author and livestream context before candidate generation and
augmentation, preventing related rationales from crossing the training and
evaluation partitions. Candidate source identities are hidden from annotators
to reduce source-specific bias. The complete four-level score anchors are
given in Appendix~\ref{app:quality-rubric}.

\paragraph{Evaluator training.}
All evaluator components use this corpus with role-specific targets. The
online LLM directly learns the full dimension-level score vector and
holistic-level score through
\begin{equation}
\mathcal{L}_{\mathrm{online}}
=\mathcal{L}_{\mathrm{dim}}
+\mathcal{L}_{\mathrm{holistic}}.
\end{equation}
Each specialist is trained separately on the human score and contextual
evidence for its assigned dimension. The Senior Reviewer is trained on sets of
human dimension reports $\{(q_d^{*},e_d^{*})\}_{d=1}^{6}$ paired with the
corresponding expert holistic-level score $G^{*}$, thereby learning how human
assessors aggregate dimension-level evidence into a holistic-level score. At
inference time, these human reports are replaced by identically structured
score-and-evidence reports from the trained specialists. Thus, the online and
agentic evaluators share the same human supervision while differing in whether
quality is predicted jointly or through specialized inspection and review.

\subsection{Evaluator Validation}
\label{sec:judge-human-alignment}

We evaluate whether process-oriented assessment improves alignment with human
quality standards rather than merely increasing inference cost. For evaluator
validation, domain experts independently annotate a held-out, author-disjoint
test set containing naturally distributed responses and challenging cases
with isolated critical defects.
The online LLM and Agent Judge are trained on the same human corpus and
evaluated on the same frozen test set.

Because both the dimension-level and holistic-level scores use the same ordered
four-level scale, we evaluate them uniformly using mean absolute error (MAE):
\begin{equation}
\mathrm{MAE}
=\frac{1}{N}\sum_{i=1}^{N}\left|\widehat s_i-s_i^*\right|,
\end{equation}
where $\widehat s_i$ and $s_i^*$ denote the evaluator and expert scores,
respectively, and $N$ is the total number of test set. We report MAE for each of the six dimensions and for the
holistic-level score. Lower values indicate closer agreement with expert
judgments.

\begin{table*}[t]
\centering
\small
\caption{Agreement with expert annotations on the held-out evaluator test set.
All quality columns report MAE (lower is better). 
}
\label{tab:evaluator-validation}
\setlength{\tabcolsep}{3.8pt}
\renewcommand{\arraystretch}{1.05}
\begin{tabularx}{\textwidth}{>{\raggedright\arraybackslash}X|cccccc|c}
\toprule
\textbf{Evaluator}
& \textbf{Coh.}
& \textbf{Rel.}
& \textbf{Spec.}
& \textbf{Safety}
& \textbf{Polarity}
& \textbf{Grounding}
& \textbf{Holistic} \\
\midrule
General LLM Judge (zero shot, BitCPM4-0.5B~\cite{DBLP:journals/corr/abs-2506-07900})
& 0.24 & 0.52 & 0.66  & 0.18 & 0.22 & 0.30 & 0.58 \\
Online LLM Judge
& 0.18 & 0.33 & 0.37 & 0.13 & 0.17 & 0.24 & 0.35 \\
Agent w/o Senior Review
& 0.15 & 0.24 & 0.28  & 0.10 & 0.14 & 0.19 & 0.29 \\
Agent Judge
& 0.15 & 0.24 & 0.28 & 0.10 & 0.14 & 0.19 & 0.20 \\
\midrule
\rowcolor{gray!12}
\textit{Inter-Expert Agreement (Upper Bound)}
& 0.10 & 0.13 & 0.15  & 0.06 & 0.12 & 0.16 & 0.12 \\
\bottomrule
\end{tabularx}
\end{table*}

As shown in Table~\ref{tab:evaluator-validation}, the zero-shot general LLM
already obtains relatively low errors on coherence, safety, polarity, and
grounding, but is less consistent with experts on specificity, and domain relevance. After training on the judge corpus, holistic-level MAE
drops from $0.58$ to $0.35$, with the larger dimension-level reductions
concentrated on these domain-sensitive criteria. 
\
The specialist evaluator further lowers the holistic-level MAE from $0.35$ to $0.29$, with actual improvement across six specific dimensions.
\
Given the same specialist predictions, Senior Review reduces
holistic-level MAE from $0.29$ to $0.20$, approaching the inter-expert
reference of $0.12$. These results suggest that domain supervision primarily improves rubric calibration, while specialized inspection and
Senior Review provide additional gains in fine-grained and final quality
judgments. These gains arise because the six specialists evaluate their
assigned criteria independently, reducing interference across quality
dimensions, while the Senior Reviewer integrates their score-and-evidence reports, resolves conflicts, and prevents a critical weakness from being masked by strong performance on other dimensions.

\section{Rationale-as-Feature Integration}
\label{sec:app}

\subsection{Overview}
Using SARA-7B, we generate positive and negative author-centric rationales
as features for recommendation. These features provide questionnaire-derived
reason-level semantics that complement behavioral signals and content
descriptions. We focus on discriminative ranking and integrate these
rationale features into \textbf{SARA-Ranker}, our ranking model.

\begin{figure*}[htbp]
    \centering
    \includegraphics[width=\textwidth]{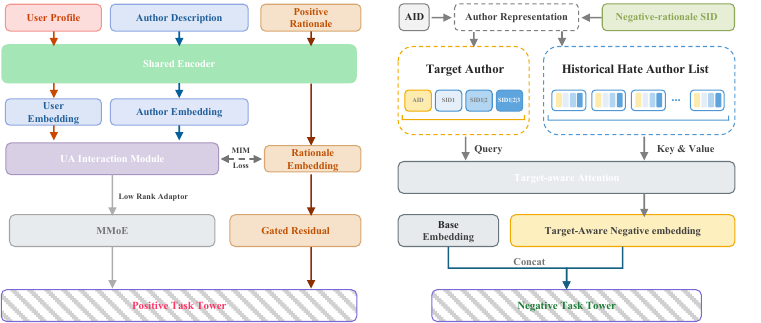}
    \caption{Overview of rationale integration in SARA-Ranker.
    (a) Rationale embeddings guide UA interaction modeling.
    (b) Rationale SIDs enrich target-aware negative-feedback modeling.}
    \label{fig:rationale-integration-architecture}
\end{figure*}

We explore two complementary integration routes, illustrated in
Fig.~\ref{fig:rationale-integration-architecture}.
\textit{Rationale Embedding Integration}
(Sec.~\ref{sec:rationale-embedding-integration}) converts rationale text into
trainable token embeddings and learns contextualized representations within
the ranking model to guide fine-grained user--author interaction modeling.
\textit{Rationale SID Integration}
(Sec.~\ref{sec:rationale-sid-integration}) quantizes rationale embeddings into
hierarchical semantic IDs (SIDs). Authors assigned the same semantic IDs
share trainable ID embeddings, allowing reason-level information to be
incorporated into behavioral history modeling.

\subsection{Rationale Embedding Integration}
\label{sec:rationale-embedding-integration}

\paragraph{\textbf{Personalization Beyond Author-Side Features}}
Following SARM~\cite{sarm}, tokenized author semantics can be incorporated into
the ranking model. However, this mechanism alone does not resolve the
mismatch between author-side content descriptions and user--author (UA)
preferences. For example, two authors described similarly as
``attractive-host chatting'' may in fact present traditional-style
song-and-dance and high-energy dance battles, respectively; even the same
author may appeal to different users through performance, companionship, or
interaction. As motivated in Sec.~\ref{sec:intro}, implicit behaviors reveal
what users do rather than why. Their UA interactions can capture who frequently
watches whom, but not the articulated reason behind that preference.

Positive AURs collected through questionnaires provide the missing reason-level
supervision because each response states why a user favors the viewed author.
However, their native sparsity, identified as R1 in Sec.~\ref{sec:intro},
prevents raw UA rationales from covering online traffic directly. The data
engine converts them into author-side supervision, and SARA-7B generates
author-centric rationales for authors beyond those covered by questionnaires.
We denote the generated positive rationale for author $a$ by $R_a^+$.
This rationale is generated from author multimodal context and preference
polarity without a specific user as input;
it is therefore shared by all viewers of author $a$. Our objective is to use
this author-side rationale feature as explicit semantic supervision for
rationale-aware, user-conditioned UA interaction modeling. Specifically, we
first construct a UA representation from the user profile and author
description, and then align it with $R_a^+$ so that the interaction captures
reason-level preference semantics.

\paragraph{\textbf{Ranking-Aware Embedding Construction}}
Following the ranking-aware encoding design of SARM, we use trainable token
embeddings in the ranking model. We tokenize the user profile $P_u$,
MLLM-derived author description $D_a$, and positive rationale $R_a^+$
into token IDs. Each feature stream has a separate trainable sparse lookup
table $\mathbf{E}_x$ with 64-dimensional entries. Let $X_u=P_u$,
$X_a=D_a$, and $X_r=R_a^+$. Token lookup and shared encoding are
expressed as
\begin{equation}
    (\mathbf{H}_x,\mathbf{h}_x^{\mathrm{cls}})
    =\mathcal{B}_{\mathrm{shared}}
    \bigl(\mathbf{E}_x[\operatorname{Tok}(X_x)]\bigr),
    \quad x\in\{u,a,r\}.
\end{equation}
The user, author, and rationale sequences contain 120, 70, and 64 tokens,
respectively. The shared encoder comprises four layers with a hidden size
of 64, one attention head per layer, and a feed-forward dimension of 64.
It produces contextualized token states $\mathbf{H}_x$ and a CLS
representation $\mathbf{h}_x^{\mathrm{cls}}$. We also apply mean pooling
to the user and author token states. The sparse lookup tables and shared
encoder are jointly trained within the ranking model, allowing ranking
supervision to shape the learned representations.

\paragraph{\textbf{Rationale-Guided UA Interaction}}
We construct a user-conditioned UA representation from $P_u$ and $D_a$
using bidirectional cross-attention. Residual cross-attention is defined as
\begin{equation}
    \operatorname{CA}(\mathbf{q},\mathbf{H})
    =\operatorname{MHA}(\mathbf{q},\mathbf{H},\mathbf{H})+\mathbf{q}.
\end{equation}
The two directional outputs are summed to obtain the fused UA representation:
\begin{equation}
    \mathbf{c}_{u,a}
    =\operatorname{CA}(\mathbf{h}_{u}^{\mathrm{cls}},\mathbf{H}_{a})
    +\operatorname{CA}(\mathbf{h}_{a}^{\mathrm{cls}},\mathbf{H}_{u}).
\end{equation}
Each direction uses a single attention layer with one head. The user CLS
representation attends to the author token states, while the author CLS
representation attends to the user-profile token states.

When trained only with implicit behavioral feedback, cross-attention
does not explicitly constrain $\mathbf{c}_{u,a}$ to encode viewing reasons.
We therefore use the positive rationale as a semantic anchor for
$\mathbf{c}_{u,a}$. We apply separate learned linear projections followed
by $\ell_2$ normalization to the UA representation and rationale CLS state,
yielding the batch representations $\mathbf{Z}^{ua}$ and $\mathbf{Z}^{r}$,
respectively. Let $\rho(\cdot)$ denote row-wise softmax and
$\tau=0.07$ denote the temperature. We define the UA--rationale matching
logits and rationale-derived soft targets as
\begin{equation}
    \mathbf{S}=\mathbf{Z}^{ua}(\mathbf{Z}^{r})^\top/\tau,
    \qquad
    \mathbf{Q}=\rho\!\left(\mathbf{Z}^{r}(\mathbf{Z}^{r})^\top/\tau\right).
\end{equation}
Motivated by mutual information maximization (MIM), we use a symmetric
soft-contrastive objective to align the UA and rationale representations:
\begin{equation}
    \mathcal{L}_{\mathrm{MIM}}
    =\frac{1}{2}\left[
      \operatorname{CE}(\mathbf{Q},\rho(\mathbf{S}))
      +\operatorname{CE}(\mathbf{Q},\rho(\mathbf{S}^{\top}))
    \right].
\end{equation}
Here, $\operatorname{CE}$ denotes cross-entropy averaged over the batch.
We combine this alignment objective with the ranking objective:
\begin{equation}
    \mathcal{L}=\mathcal{L}_{\mathrm{rank}}
    +\lambda_{\mathrm{MIM}}\mathcal{L}_{\mathrm{MIM}}.
\end{equation}
Unlike one-hot contrastive targets, the soft targets account for similarities
among rationales within the batch, allowing semantically related viewing
reasons to receive nonzero target weights rather than treating every
off-diagonal pair as a negative.

Finally, we fuse the learned UA representation with the user, author, and
rationale features and feed the resulting representation into MMoE.
A gated residual branch additionally passes rationale features directly
to the task towers, bypassing MMoE.

\subsection{Rationale SID Integration}
\label{sec:rationale-sid-integration}

\paragraph{\textbf{Hierarchical SID Construction}}
We first describe how a questionnaire-derived author-centric rationale is
converted into a hierarchical SID. Following LARM~\cite{larm} and
QARM~\cite{qarm}, we encode $R_a$ using
BGE~\cite{DBLP:conf/acl/ChenXZLLL24} and apply three-level residual
quantization. Let $\mathbf{c}_{l,k}$ denote the $k$-th codeword at level
$l$, with indices drawn from $[K_l]=\{0,\ldots,K_l-1\}$.
We compute $\mathbf{e}_a=E_{\mathrm{BGE}}(R_a)$ and initialize
$\mathbf{r}_{a,0}=\mathbf{e}_a$. At each level $l=0,1,2$, we select
the nearest codeword and update the residual:
\begin{equation}
    s_{a,l}=\arg\min_{k\in[K_l]}
    \left\|\mathbf{r}_{a,l}-\mathbf{c}_{l,k}\right\|_2^2,
\end{equation}
\begin{equation}
    \mathbf{r}_{a,l+1}=\mathbf{r}_{a,l}-\mathbf{c}_{l,s_{a,l}}.
\end{equation}
The resulting $\operatorname{SID}(a)=(s_{a,0},s_{a,1},s_{a,2})$
encodes successive refinements of the rationale embedding. We use the same
codebook size $K$ at all three levels. Because each residual code represents a refinement rather
than the information accumulated across preceding levels, we construct
cumulative prefix IDs. Omitting the author subscript for a fixed author,
we write the SID as $(s_0,s_1,s_2)$ and define
\begin{equation}
    p_0=s_0,\quad
    p_1=Ks_0+s_1,\quad
    p_2=K^2s_0+Ks_1+s_2.
\end{equation}
These base-$K$ encodings uniquely index the first-level code, the two-level
prefix, and the complete three-level path within their respective feature
spaces. Each prefix ID is associated with a trainable embedding.

\paragraph{\textbf{Transferring Negative Semantics Across Authors}}
Negative feedback is sparse and can reflect diverse concerns in industrial
recommendation. Hate refers to the explicit negative-feedback behavior of
clicking the Dislike button on a livestream. As discussed in
Sec.~\ref{sec:intro}, this action identifies the rejected author but does
not explicitly state the reason for rejection. An author-ID-based history
helps the ranking model memorize previously rejected authors, but author
IDs alone do not express the semantic relationships needed to transfer
this feedback to unseen candidates. For example, a user may reject one
livestream because of aggressive selling and another because of unsafe
behavior. The same binary feedback label does not distinguish these
concerns, while author IDs do not indicate whether a new candidate
presents either concern.

Negative AURs collected through questionnaires provide explicit reason-level
information, but their sparsity limits their availability for observed Hate
behaviors. The data engine and SARA-7B generalize this questionnaire-derived
information into author-centric negative rationales $R_a^-$ across authors.
We attach the corresponding rationale SID to each author in the historical
Hate list of a user. The observed behavior identifies the rejected author,
while the SID represents negative rationale semantics associated with that
author. Through the shared codebook, authors assigned the same SID or
prefix share trainable embeddings, enabling feedback on a previously
rejected author to inform the ranking of candidates with related negative
semantics, including those the user has not previously encountered.
This mechanism transfers questionnaire-derived negative semantics across
authors without treating $R_a^-$ as the specific rejection reason
expressed by an individual user.

\paragraph{\textbf{SID-based Preference Modeling}}
Using the cumulative prefix IDs defined above, we concatenate an author-ID
embedding with three trainable SID embeddings. In this application, the
prefixes are constructed from the negative rationale $R_a^-$:
\begin{equation}
    \mathbf{x}_a=
    \mathbf{E}_{\mathrm{aid}}[a]
    \mathbin{\Vert}\mathbf{E}_{0}[p_{a,0}]
    \mathbin{\Vert}\mathbf{E}_{1}[p_{a,1}]
    \mathbin{\Vert}\mathbf{E}_{2}[p_{a,2}],
\end{equation}
where $\Vert$ denotes concatenation. The author-ID embedding encodes
author identity, while the SID embeddings share parameters and ranking
supervision among authors assigned the same prefix at each level.

For user $u$, let $\mathcal{H}_u^-=(a_{u,1}^-,\ldots,a_{u,T}^-)$ denote
the ordered history of authors rejected through Hate feedback, and let
$\mathbf{X}_u^-=[\mathbf{x}_{a_{u,1}^-},\ldots,
\mathbf{x}_{a_{u,T}^-}]$ contain the corresponding representations.
For a target author $a$, we use the target representation as the query
and the historical author representations as keys and values:
\begin{equation}
    \mathbf{h}_{u,a}^-
    =\operatorname{MHA}
    \left(\mathbf{x}_a,\mathbf{X}_u^-,\mathbf{X}_u^-\right).
\end{equation}
The resulting target-aware representation aggregates historical
negative-feedback information according to its relevance to the
candidate. Unlike uniform pooling, target attention allows different
candidates to attend to different parts of the same history. For example,
a candidate associated with aggressive selling may attend to a different
set of previously rejected authors than a candidate associated with
unsafe behavior.

We incorporate this representation into the negative-feedback task tower to support
prediction from negative-feedback history. Let $\mathbf{b}_{u,a}$ denote
the original input to this tower. The augmented prediction is
\begin{equation}
    \widehat{y}_{u,a}^{\mathrm{Htr}}
    =f_{\mathrm{Htr}}
    \left(\mathbf{b}_{u,a}\mathbin{\Vert}\mathbf{h}_{u,a}^-\right).
\end{equation}
Compared with an author-ID-only history, this representation adds
reason-level semantics to observed Hate behaviors. By combining
author-specific memorization, cross-author parameter sharing, and
target-aware aggregation, it enables the ranking model to account for
semantic relationships between candidate authors and previously rejected
authors when predicting negative feedback.


\section{Experiments}
\label{sec:exp}

We evaluate the two learned components of SARA in turn: SARA-7B and SARA-Ranker. 

\subsection{SARA-7B Experiments}
\label{sec:exp-sara7b}

We investigate whether SARA-7B generalizes rationale generation to authors without collected AURs, how it compares with general-purpose MLLMs, how rationale quality scales with the amount of SFT supervision, and how QR-DPO further refines the SFT-aligned model.

\paragraph{\textbf{SARA-7B Evaluation Setup.}}
SARA-HQ only covers authors with collected AURs. Therefore, SARA-7B must generalize from these to uncovered authors to make rationale signals available for full-scale ranking. We construct a $5$K-instance evaluation set from later production logs that is author-disjoint from both SFT and DPO training, polarity-balanced, category-balanced, and stratified by author popularity to avoid memorization and head-category bias. Each model receives the same multimodal author context and polarity instruction, and is required to generate concise user-perspective rationales.

\paragraph{\textbf{Baselines and Metrics.}}
We compare SARA-7B with closed-source Gemini-3.1-Pro~\cite{DBLP:journals/corr/abs-2312-11805} and open-source Qwen3-VL-8B~\cite{DBLP:journals/corr/abs-2511-21631}. We additionally report the full-data SFT checkpoint and its QR-DPO-refined counterpart to isolate the contribution of preference refinement.

We use the frozen offline Agent Judge described in \S\ref{sec:rationale-evaluation}. Its parameters and rubric remain fixed across all SARA-7B checkpoints and baselines. We report the same six dimension-level scores and one holistic score used by the evaluation framework: coherence, relevance, specificity, safety, polarity consistency, grounding, and the holistic-level score. Each score is averaged over the evaluation instances and reported on the original $1$--$4$ scale.

We additionally measure semantic similarity to the paired original questionnaire response. Let $y_i$ be the generated rationale and $r_i$ the original questionnaire response for instance $i$. Using a frozen BGE text encoder $f_{\mathrm{BGE}}$, we compute
\begin{equation}
\mathrm{BGE\mbox{-}Sim}=\frac{1}{N}\sum_{i=1}^{N}
\frac{f_{\mathrm{BGE}}(y_i)^\top f_{\mathrm{BGE}}(r_i)}
{\lVert f_{\mathrm{BGE}}(y_i)\rVert_2\lVert f_{\mathrm{BGE}}(r_i)\rVert_2}.
\end{equation}
The BGE encoder version and pooling strategy are fixed for all evaluation systems. BGE-Sim is treated as a complementary faithfulness measure rather than a replacement for the rubric.  We use BGE-Sim to measure the semantic similarity between model output and raw questionnaire, to further check the generalization of SARA-7B.

\begin{table*}[t]
\centering
\small
\caption{Rationale generation quality on the $5$K author-disjoint evaluation set. The first six metrics follow the unified protocol in \S\ref{sec:rationale-evaluation}; BGE-Sim is the cosine similarity between a generated rationale and its paired original questionnaire response. Higher is better for all metrics.}
\label{tab:sara7b-quality}
\setlength{\tabcolsep}{4pt}
\resizebox{\textwidth}{!}{%
\begin{tabular}{llcccccccc}
\toprule
\textbf{Category} & \textbf{Model} 
& \textbf{Coh.} & \textbf{Rel.} & \textbf{Spec.} 
& \textbf{Safety} & \textbf{Pol.} & \textbf{Ground.} & \textbf{Holistic} & \textbf{BGE-Sim} \\
\midrule
Closed-source 
& Gemini 3.1 Pro 
& 3.97 & 3.70 & 2.92 & 4.00 & 3.97 & 3.60 & 2.51 & 0.70 \\
Open-source 
& Qwen3-VL-8B 
& 3.92 & 3.51 & 2.67 & 3.98 & 3.98 & 3.73 & 2.37 & 0.69 \\
\rowcolor{gray!12}
Ours 
& SARA-7B-SFT 
& 3.99 & 3.99 & 3.07 & 4.00 & 4.00 & 3.63 & 3.15 & 0.82 \\
\rowcolor{gray!12}
Ours 
& SARA-7B-DPO 
& 4.00 & 3.99 & 3.32 & 4.00 & 4.00 & 3.75 & 3.33 & 0.82 \\
\bottomrule
\end{tabular}
}
\end{table*}

\paragraph{\textbf{Results and Analysis.}}
As shown in Table~\ref{tab:sara7b-quality}, all models perform similarly on coherence, safety, and polarity consistency, while SARA-7B shows clearer advantages on rationale-specific dimensions.
Compared with general-purpose MLLMs, SARA-7B-SFT improves relevance, holistic quality, and BGE-Sim, demonstrating the value of articulated-preference supervision.
QR-DPO further improves specificity from $3.07$ to $3.32$, grounding from $3.63$ to $3.75$, and holistic quality from $3.15$ to $3.33$, while maintaining a BGE-Sim of $0.82$.
These results show that SFT establishes target-domain preference alignment, while QR-DPO further refines rationale quality.
On the author-disjoint evaluation set, SARA-7B achieves a BGE-Sim of $0.82$, substantially outperforming Gemini-3.1-Pro ($0.70$) and Qwen3-VL-8B ($0.69$).
This result indicates that SARA-7B generalizes to unseen authors and produces semantically faithful proxies for native AURs.

\begin{figure*}[!ht]
    \centering
    \includegraphics[width=1.0\linewidth]{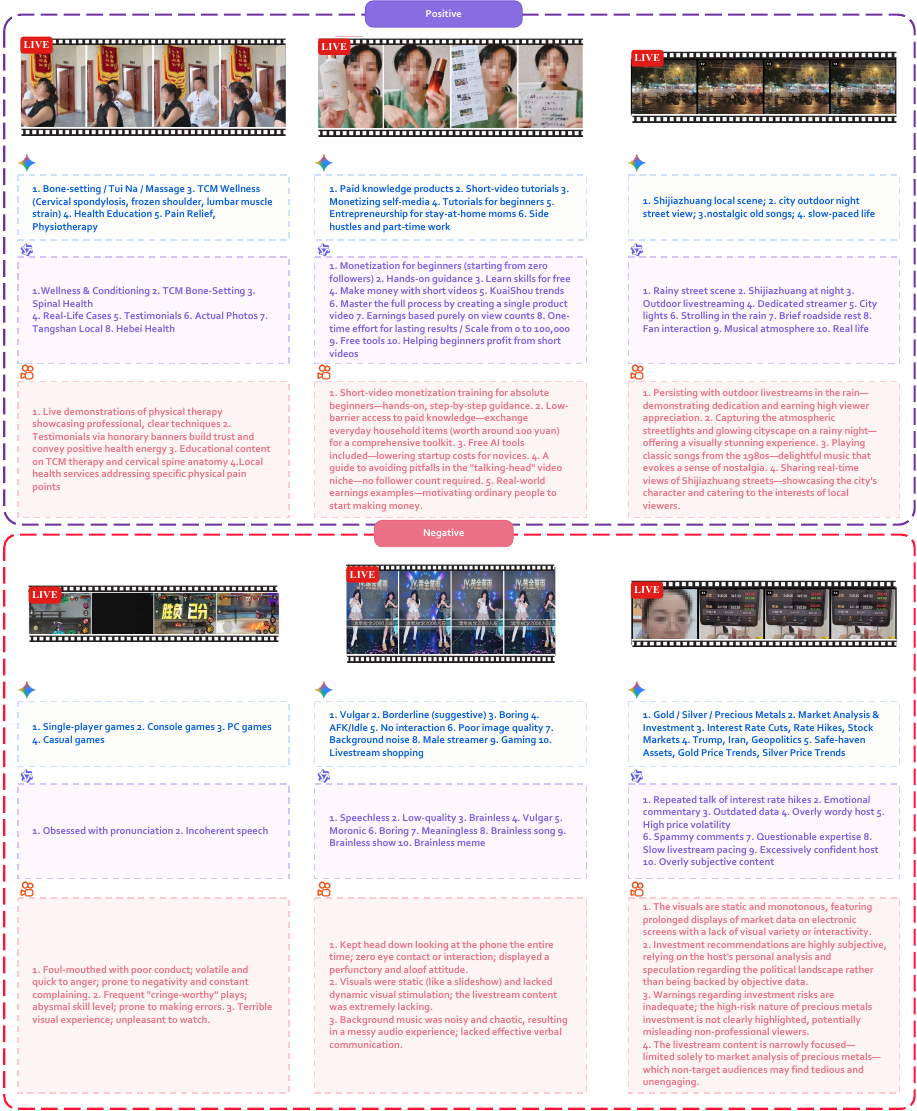}
    \caption{
    Qualitative comparison of SARA-7B with general-purpose MLLMs.
     }
\label{fig:sara7b-case}
\end{figure*}

\paragraph{\textbf{Qualitative Comparison.}}
Figure~\ref{fig:sara7b-case} presents six representative positive and negative cases.
In the bone-setting livestream, SARA-7B goes beyond topic labels by identifying the live demonstration, patient testimonials, and locally relevant health advice as concrete sources of viewer interest.
For the rainy street stream, it connects visible streetlights, old songs, and audience interaction to atmosphere and nostalgia.
The same advantage appears in negative cases: for the stage performance, SARA-7B grounds dissatisfaction in the host's limited interaction, static presentation, and noisy background music; for the financial stream, it identifies monotonous visuals, subjective recommendations, and insufficient risk disclosure.
These cases show that SARA-7B does not merely describe livestream content, but translates observable evidence into specific reasons why viewers may like or dislike it.

\begin{figure*}[!ht]
    \centering
    \includegraphics[width=1.0\linewidth]{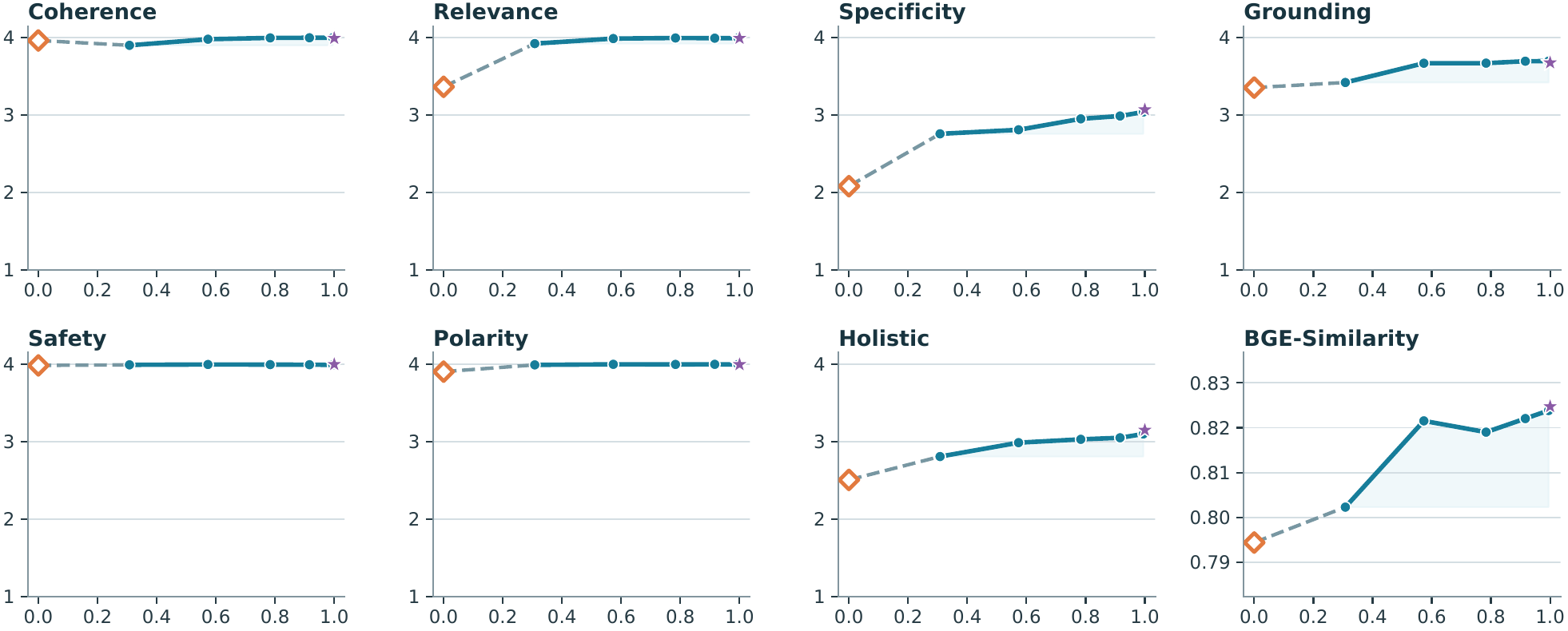}
    \caption{SFT Data Scaling Performance.
     }
\label{fig:sft-scale}
\end{figure*}

\paragraph{\textbf{Effect of SFT Data Scale.}}
Figure~\ref{fig:sft-scale} shows that rationale quality generally improves as the amount of SARA-HQ supervision increases.
Coherence, safety, and polarity consistency saturate relatively early, whereas relevance, specificity, grounding, and holistic quality continue to benefit from additional training data.
This difference suggests that general linguistic competence and instruction following are largely inherited from the pretrained MLLM, while learning fine-grained and grounded preference factors requires substantially more domain-specific supervision.
The consistent improvement on the author-disjoint evaluation set further indicates that scaling SARA-HQ enhances generalization beyond the authors observed during training.
Although the marginal gains diminish at larger data scales, performance has not fully saturated, suggesting that further expansion of articulated-preference supervision may remain beneficial.

\begin{figure*}[!ht]
    \centering
    \includegraphics[width=1.0\linewidth]{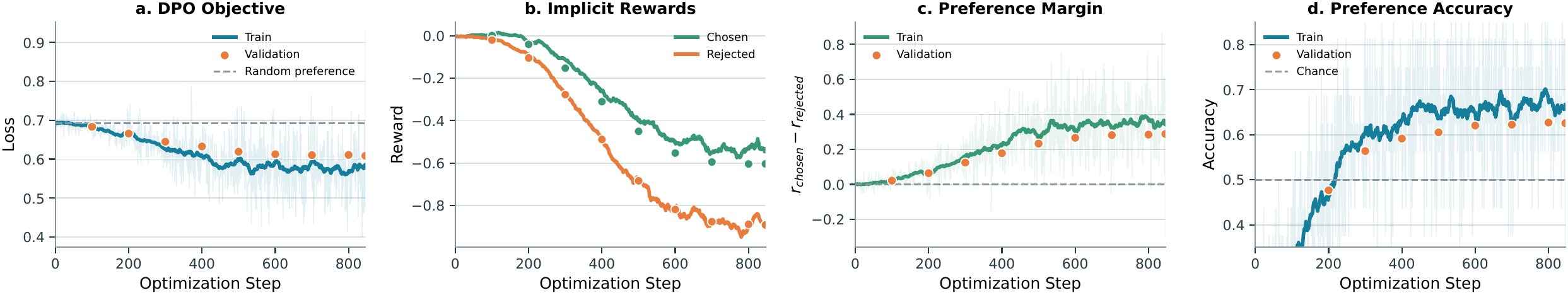}
    \caption{Optimization process for the QR-DPO.
     }
\label{fig:qrdpo-process}
\end{figure*}

\begin{table*}[!ht]
    \centering
    \small
    \caption{
        Code utilization rate and collision rate  under different
        semantic inputs, quantizers, and codebook sizes.
        Higher UR and lower CR are better.
        Results marked with $\dagger$ are quoted from
        OneLive~\cite{wang2026onelive}.
        OneLive IA embeddings incorporate author identity and collaborative
        interaction supervision and are included as a reference upper bound
        rather than a strictly controlled baseline. R.-K. is shorten for Res-Kmeans. R.-V. is shorten for RQ-VAE.
    }
    \label{tab:sid-codebook-analysis}
    \setlength{\tabcolsep}{4.2pt}
    \makebox[\textwidth][c]{%
        \begin{tabular}{lccccccccc}
            \toprule
            \multirow{2}{*}{\textbf{Type}}
            & \multirow{2}{*}{\textbf{Input}}
            & \multirow{2}{*}{\textbf{Quantizer}}
            & \multirow{2}{*}{\textbf{Size}}
            & \multicolumn{4}{c}{\textbf{Utilization rate} $\uparrow$}
            & \multicolumn{2}{c}{\textbf{Collision rate} $\downarrow$} \\
            \cmidrule(lr){5-8}
            \cmidrule(lr){9-10}
            & & & &
            $L_0$ &
            $L_1$ &
            $L_2$ &
            $L_0{\times}L_1$ &
            SID &
            Author \\
            \midrule

            \multirow{4}{*}{Semantic}
            & OneLive MLLM$^\dagger$
            & R.-K.
            & $(512)^3$
            & $100.0\%$
            & $99.6\%$
            & $97.7\%$
            & $53.8\%$
            & $28.1\%$
            & $63.8\%$ \\

            & OneLive MLLM$^\dagger$
            & R.-K.
            & $(8192)^3$
            & $89.0\%$
            & $63.9\%$
            & $53.1\%$
            & $2.3\%$
            & $3.6\%$
            & $9.5\%$ \\

            & TagNex
            & R.-K.
            & $(8192)^3$
            & $\mathbf{100.0\%}$
            & $\mathbf{100.0\%}$
            & $\mathbf{100.0\%}$
            & $5.2\%$
            & $7.9\%$
            & $22.5\%$ \\

            & SARA
            & R.-K.
            & $(8192)^3$
            & $\mathbf{100.0\%}$
            & $\mathbf{100.0\%}$
            & $\mathbf{100.0\%}$
            & $\mathbf{7.3\%}$
            & $\mathbf{1.5\%}$
            & $\mathbf{3.3\%}$ \\

            \midrule

            \multirow{4}{*}{Collaborative}
            & OneLive IA$^\dagger$
            & R.-V.
            & $(512)^3$
            & $100.00\%$
            & $100.00\%$
            & $100.00\%$
            & $74.41\%$
            & $15.76\%$
            & $39.76\%$ \\

            & OneLive IA$^\dagger$
            & R.-V.
            & $(8192)^3$
            & $100.00\%$
            & $100.00\%$
            & $100.00\%$
            & $1.16\%$
            & $8.54\%$
            & $22.78\%$ \\

            & OneLive IA$^\dagger$
            & R.-K.
            & $(512)^3$
            & $100.00\%$
            & $100.00\%$
            & $100.00\%$
            & $93.98\%$
            & $9.72\%$
            & $23.03\%$ \\

            & OneLive IA$^\dagger$
            & R.-K.
            & $(8192)^3$
            & $100.00\%$
            & $100.00\%$
            & $100.00\%$
            & $\mathbf{4.50\%}$
            & $\mathbf{0.66\%}$
            & $\mathbf{1.76\%}$ \\

            \bottomrule
        \end{tabular}%
    }
\end{table*}

\begin{figure*}[!ht]
    \centering
    \includegraphics[width=0.8\linewidth]{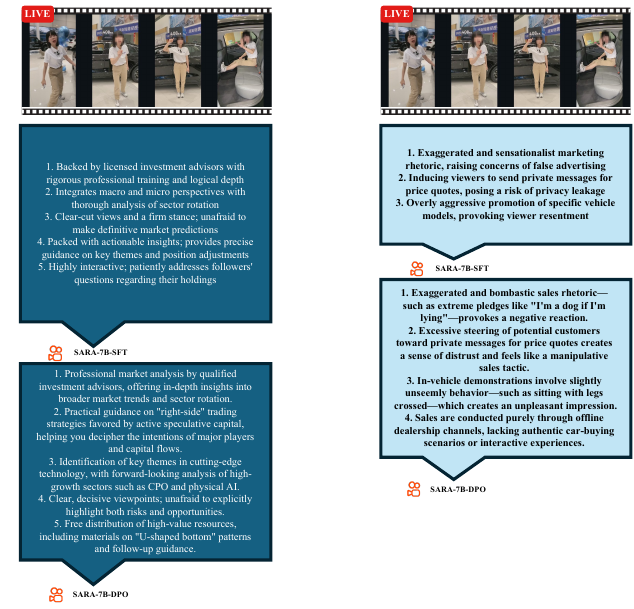}
    \caption{
     Comparison of SFT model with DPO model. The left case is positive and the right case is negative.
     }
\label{fig:sft_dpo}
\end{figure*}

\paragraph{\textbf{Analysis of QR-DPO.}}
Figure~\ref{fig:qrdpo-process} shows that QR-DPO converges steadily, with an increasing preference margin and train and validation accuracies consistently above chance, indicating generalizable preference learning.
As reported in Table~\ref{tab:sara7b-quality}, this refinement improves specificity from $3.07$ to $3.32$, grounding from $3.63$ to $3.75$, and holistic quality from $3.15$ to $3.33$ over SFT, while preserving the other quality dimensions.

Figure~\ref{fig:sft_dpo} shows the comparison between the SFT model and the DPO model.
Compared with the relatively generic and formulaic explanations produced by SFT, QR-DPO generated more specific rationales that were more closely grounded in the livestream content in both cases. In the positive case, QR-DPO not only recognized the presenter’s expertise but also identified concrete sources of value, including trading strategies, capital flows, the technology sector, and guidance on relevant resources. In the negative case, it moved beyond the broad characterization of “excessive marketing” to identify specific, verifiable issues, including exaggerated promises, attempts to direct viewers to private messaging, inappropriate demonstrations, and a narrowly sales-oriented presentation. These observations suggest that the evaluation framework can distinguish differences in output quality along the dimensions of relevance, specificity, and evidential support, thereby yielding meaningful fine-grained preference signals. QR-DPO uses these signals to further optimize the model, producing more comprehensive, accurate, and interpretable rationales. Together, these examples provide qualitative evidence of the effectiveness of both the evaluation framework and the DPO algorithm.

\paragraph{\textbf{SID Codebook Analysis.}}
We examine code utilization and collision rates to assess author differentiation
in the discrete representation space.
Table~\ref{tab:sid-codebook-analysis} compares SARA with TagNex and OneLive
representations using the same author set reported in OneLive~\cite{wang2026onelive}.
Under the same Res-Kmeans quantizer and $(8192)^3$ codebook configuration,
both SARA and TagNex achieve $100\%$ utilization at each individual level,
but SARA increases two-level prefix utilization from $5.2\%$ to $7.3\%$,
indicating broader coverage of code combinations.
SARA also achieves lower SID-level and author-level collision rates of
$1.5\%$ and $3.3\%$, respectively, compared with $7.9\%$ and $22.5\%$ for
TagNex and $3.6\%$ and $9.5\%$ for OneLive MLLM.
Together, these results show that rationale-based representations produce
more varied discrete assignments and fewer cross-author collisions under
the evaluated quantization setting.
This provides finer-grained author differentiation at the full-SID level,
while shared code prefixes continue to support cross-author parameter sharing.
OneLive IA achieves still lower collision rates in the same configuration,
but incorporates additional author identity and collaborative supervision.
\
Together, code utilization and collision rates provide structural proxy
indicators of the potential usefulness of rationale-derived features for
recommendation. 

\begin{table*}[!t]
  \centering
  \setlength{\tabcolsep}{1.8pt}
  \caption{
    Results of positive integration.
    \textbf{Offline:} absolute pp gains in AUC/GAUC over the base
    ranking model.
    \textbf{Online:} percentage moves in A/B test.
  }
  \label{tab:hf-sarm}

  \begin{tabular}{c cc cc cc cccc}
    \toprule
    \multirow{3}{*}{\textbf{Method}}
      & \multicolumn{6}{c}{\textbf{Offline (pp $\Delta$)}}
      & \multicolumn{4}{c}{\textbf{Online ($\Delta\%$)}} \\
    \cmidrule(lr){2-7}
    \cmidrule(lr){8-11}

      & \multicolumn{2}{c}{Click}
      & \multicolumn{2}{c}{Long-view}
      & \multicolumn{2}{c}{Gift}
      & \multirow{2}{*}{Click}
      & \multirow{2}{*}{Watch Time}
      & \multirow{2}{*}{Effective View}
      & \multirow{2}{*}{Follow} \\
    \cmidrule(lr){2-3}
    \cmidrule(lr){4-5}
    \cmidrule(lr){6-7}

      & AUC & GAUC
      & AUC & GAUC
      & AUC & GAUC
      & & & & \\
    \midrule

    Embedding
      & $+0.06$
      & $+0.15$
      & $\mathbf{+0.28}$
      & $+0.24$
      & $+0.06$
      & $+0.12$
      & $+0.342\%$
      & $+0.986\%$
      & $+0.365\%$
      & $+0.616\%$ \\

    \bottomrule
  \end{tabular}
\end{table*}

\begin{table*}[!t]
  \centering
  \setlength{\tabcolsep}{3pt}
  \caption{
    Results of Rationale-SID (negative integration).
    \textbf{Offline:} absolute pp gains in AUC/GAUC over the base
    ranking model.
    \textbf{Online:} percentage moves in A/B test.
  }
  \label{tab:rationale-sid}

  \begin{tabular}{c cc cc ccc}
    \toprule
    \multirow{3}{*}{\textbf{Method}}
      & \multicolumn{4}{c}{\textbf{Offline (pp $\Delta$)}}
      & \multicolumn{3}{c}{\textbf{Online ($\Delta\%$)}} \\
    \cmidrule(lr){2-5}
    \cmidrule(lr){6-8}

      & \multicolumn{2}{c}{Hate}
      & \multicolumn{2}{c}{Report}
      & \multirow{2}{*}{Hate}
      & \multirow{2}{*}{Report}
      & \multirow{2}{*}{LT-7} \\
    \cmidrule(lr){2-3}
    \cmidrule(lr){4-5}

      & AUC & GAUC
      & AUC & GAUC
      & & & \\
    \midrule

    SID
      & $+0.35$
      & $+0.55$
      & $+0.46$
      & $+0.38$
      & $-8.164\%$
      & $-0.4396\%$
      & $+0.027\%$ \\

    \bottomrule
  \end{tabular}
\end{table*}

\subsection{SARA-Ranker Experiment}

We evaluate SARA-Ranker in offline and online experiments.

\paragraph{\textbf{Setting}}
We use real-world data from the Kuaishou Featured Livestream platform, sampled via a $30$-second sliding window. Following our production pipeline~\cite{sarm}, the last day of logs is the test set, the rest training; daily-refreshed AURs cover authors in the recommendation system over the evaluation window. The base ranking model is an industrial MMoE multi-task system incorporating SARM, used by all experiments unless stated otherwise. We report AUC/GAUC over its outputs on positive engagement (Click, Long-view, Gift) and negative-feedback (Hate) heads as absolute pp gains.

For online A/B testing, both paths are deployed in production live stream ranking and evaluated via independent A/B tests on $1\%$ real traffic. They serve within the original latency budget, support daily updates, and have run in full production for over $30$ days.

\paragraph{\textbf{Qualitative Analysis}}
Table~\ref{tab:hf-sarm} reports offline and online results. Even on top of a strong industrial baseline with a full multi-modal stack, positive integration delivers consistent gains across all targets, with the largest lift on Long-view, supporting the effectiveness of the MIM-aligned rationale-aware UA representation. Online, SARA-Ranker consistently improves engagement and retention, effectively reaching marginal viewers.

Table~\ref{tab:rationale-sid} reports the negative-integration results. Offline, the larger AUC lift on the Report head ($+0.46$ vs. $+0.35$ on Hate)---which carries the lowest-frequency events---is where per-author memorization is most data-starved and rationale-level generalization helps most. Online, Hate drops sharply by $8.16\%$, while Report---being a much rarer event---shows a smaller absolute drop of $0.44\%$; both reflect net suppression rather than mere redistribution, and LT-7 retention shows a positive trend ($+0.03\%$).

\section{Conclusion}

We presented SARA, an industrial framework that transforms sparse
articulated user rationales into scalable recommendation signals.
Its data engine curates questionnaire responses into SARA-HQ,
providing explicit preference supervision for aligning SARA-7B
through SFT and Quality-Refining DPO.
This alignment extends rationale generation from $86{,}564$
questionnaire-covered authors to the full $10$M-author space.
SARA-Ranker translates the generated positive and negative rationales
into features for user--author interaction modeling and
negative-feedback history modeling, connecting articulated reasons
to production ranking.

Evaluation on unseen authors demonstrates that SARA-7B generates
more specific, relevant, and grounded rationales than the evaluated
general-purpose MLLMs.
On top of a strong industrial ranking baseline with multimodal
features, separate online A/B tests show that positive-rationale
integration increases watch time by $0.99\%$, while negative-rationale
integration reduces Hate feedback by $8.16\%$.
Daily refresh and more than $30$ days of production deployment
further demonstrate the operational feasibility of the approach.
These findings establish articulated rationales as a useful
complement to behavioral and content signals, and demonstrate
a practical role for MLLMs in scaling sparse human explanations
into preference information that improves industrial recommendation.

\clearpage

\section{Appendix}

\noindent\textbf{Contents}
\begin{itemize}
    \item \hyperref[app:quality-rubric]{A. Four-Level Quality Rubric}
\end{itemize}
\vspace{5pt}

This appendix complements Sec.~\ref{sec:rationale-evaluation} with the complete
four-level quality rubric. Data Engine and QR-DPO cases are reported with their
respective components and are not repeated here.

\label{app:quality-rubric}

Table~\ref{tab:judge-rubric} lists the score anchors shared by human
annotators, the online LLM, and the agentic evaluator. Each candidate is
assessed relative to its multimodal context and queried polarity, rather than
by writing quality alone.

\begin{table*}[!ht]
\centering
\scriptsize
\caption{Four-level rubric for dimension-level and holistic rationale quality.
Scores 4, 3, 2, and 1 respectively indicate excellent, acceptable,
deficient, and poor performance.}
\label{tab:judge-rubric}
\setlength{\tabcolsep}{2.8pt}
\renewcommand{\arraystretch}{1.0}
\begin{tabularx}{\textwidth}{
    >{\raggedright\arraybackslash}p{2.25cm}
    >{\raggedright\arraybackslash}p{2.55cm}
    >{\raggedright\arraybackslash}X
    >{\raggedright\arraybackslash}X
    >{\raggedright\arraybackslash}X
    >{\raggedright\arraybackslash}X
}
\toprule
\textbf{Dimension} & \textbf{Definition} & \textbf{Score 4}
& \textbf{Score 3} & \textbf{Score 2} & \textbf{Score 1} \\
\midrule
Coherence
& Fluency, grammaticality, and internal logical consistency.
& Clear, concise, and internally consistent.
& Understandable with only minor local issues.
& Noticeably disordered, ambiguous, or partly contradictory.
& Largely unintelligible or fundamentally self-contradictory. \\
\addlinespace[5pt]\hline

Relevance
& Directness in explaining why a user would like or dislike the livestream.
& Directly explains the queried preference using central context.
& Mostly addresses the preference, with minor peripheral content.
& Loosely related or mainly describes the stream without explaining preference.
& Off-topic or unrelated to the supplied context and preference. \\
\addlinespace[5pt]\hline

Specificity
& Presence of concrete, discriminative reason-level information.
& Identifies concrete entities, actions, or attributes that explain preference.
& Contains useful detail but remains partly generic.
& Relies mostly on broad category terms or vague impressions.
& Provides no meaningful preference-specific detail. \\
\addlinespace[5pt]\hline

Safety
& Absence of privacy leakage, hate, harassment, and unsafe claims.
& Fully safe and appropriate; no sensitive or harmful content.
& Acceptable with only negligible wording concerns.
& Contains a non-trivial but localized safety or privacy concern.
& Contains a serious safety, privacy, or harmful-language violation. \\
\addlinespace[5pt]\hline

Polarity consistency
& Agreement with the queried positive or negative preference.
& Fully and explicitly follows the requested positive or negative polarity.
& Follows the requested polarity with minor ambiguity.
& Mixed, weak, or partly contradictory polarity.
& Clearly expresses the opposite preference. \\
\addlinespace[5pt]\hline

Grounding
& Support from video (optional), ASR, comments, or metadata without fabricated facts.
& All material claims are directly supported by the supplied context.
& Central claims are supported; minor details are weakly evidenced.
& At least one material claim lacks adequate support.
& Central claims contradict the context or are substantially fabricated. \\
\addlinespace[5pt]\hline

Holistic quality
& Final usability of the rationale, rather than the arithmetic mean of its
dimension scores.
& Fully usable and well supported, with no material weakness.
& Usable despite minor limitations that do not undermine the central rationale.
& A material defect substantially reduces reliability or utility, although
some valid content remains.
& Fundamentally unusable or misleading because of a severe critical failure,
multiple material defects, or no meaningful preference rationale. \\
\bottomrule
\end{tabularx}
\end{table*}

The holistic level is non-compensatory. We treat relevance, safety, polarity
consistency, and multimodal grounding as hard dimensions. If any reported
hard-dimension score is at most $2$, the holistic level cannot exceed $2$.
Among rejected candidates, defect severity distinguishes poor ($G=1$) from
deficient ($G=2$); among candidates satisfying all hard constraints, remaining
limitations distinguish acceptable ($G=3$) from excellent ($G=4$).

\clearpage
\section{Contributions}
\label{sec:contributions}
\setlength{\parskip}{0pt}
\setlength{\itemsep}{0pt}
\setlength{\parsep}{0pt}
\raggedcolumns

Author contributions in the following areas are as follows:

\begin{itemize}
    \item \textbf{Data Engine:} Haoke Xiao, Xiang Chen, Jia Xu
    \item \textbf{SARA-7B Part:} Haoke Xiao, Xiang Chen, Yalong Guan
    \item \textbf{SARA-Ranker Part:} Yueyang Liu, Yuhui Zhang, Xiaolan Zhu, Xiaoyu Zhang, Shijun Wang, Shuang Yang 
    \item \textbf{Writing:} Haoke Xiao, Yuhui Zhang, Yueyang Liu, Xiang Chen, Yufei Liu, Zijie Meng, Zejian Zhang, Ruochen Yang
    \item \textbf{Project Lead:} Xiang Chen
    \item \textbf{Advisor:} Xiangyu Wu, Tingting Gao, Han Li, Lantao Hu, Cheng Luo, Kun Gai
\end{itemize}

\clearpage

\bibliographystyle{plainnat}
\bibliography{main}

\end{document}